\documentclass[preprint,amsmath,amssymb,prb,showkeys]{revtex4}

\newcommand{\mbf}[1]{\mbox{\boldmath $#1$}}

\def\0{\mbox{\mbf 0}}

\def\BibTeX{{\rm B\kern-.05em{\sc i\kern-.025em b}\kern-.08em
    T\kern-.1667em\lower.7ex\hbox{E}\kern-.125emX}}

\usepackage{graphicx}

\usepackage{dcolumn}

\usepackage{bm}

\begin{document}


\title{A `Cellular Neuronal' Approach to Optimization Problems\\(2nd revised)}

\author{Gregory S. Duane\footnote{Corresponding Author: gregory.duane@colorado.edu 
303-492-7263  fax: 303-492-3524}}
\affiliation{%
National Center for Atmospheric Research, P.O. Box 3000, Boulder, CO
80307\\and\\
School of Mathematics, University of Minnesota, 206 Church St. S.E., Minneapolis, MN 55455 }

%

\begin{abstract}
The Hopfield-Tank (1985) recurrent neural network architecture for the Traveling
Salesman Problem is generalized to a fully interconnected ``cellular'' neural network of
regular oscillators.  Tours are defined by synchronization patterns, allowing 
the simultaneous representation of all cyclic permutations of a given tour. 
The network converges to local optima some of which correspond to
shortest-distance tours, as can be shown analytically in a stationary phase
approximation. Simulated
annealing is required for global optimization, but the stochastic element might
be replaced by chaotic intermittency in a further generalization of
the architecture to a network of chaotic oscillators.

\end{abstract}

%



\maketitle

{\bf An artificial neural network with a symmetric connection allowed between every pair of
``neuronal" units evolves to a state that satisfies an optimization principle expressed in terms of the
weights defining those connections. In a Hopfield associative memory network\cite{Hop82}, the optimization
principle is satisfied when the network settles to one of many pre-stored patterns, depending
on the initial state. In the Hopfield-Tank\cite{HopTank} construction, weights were chosen so
that the stable fixed points of the network correspond to shortest-distance tours among
a set of cities arbitrarily laid out on a map, offering a potential solution to the algorithmically difficult
Travelling Salesman Problem (TSP). A stochastic element was introduced in a ``simulated
annealing" scheme so that the network would escape local optima, such as those corresponding
to good tours that are not necessarily shortest. Here, the Hopfield-Tank scheme is
generalized to a ``cellular neural network" in which the units are dynamical systems that
oscillate in the absence of input from other units, arguably resembling real neurons
a bit more closely. The representation of the TSP turns out to be somewhat more natural in
such a network, where tours are defined in terms of synchronization patterns among
the oscillating units.  Additionally, intermittent synchronization
might naturally fill the role of the stochastic element in the original scheme.  
In this work we construct a network of regular oscillators 
that may solve the TSP about as well as the original Hopfield-Tank network, as preparation
for an extension to a network of chaotic oscillators.}

\section{Introduction}

The Hopfield-Tank neural network solution to the 
Travelling Salesman Problem was a milestone in neuromorphic computation.
While the proposed architecture provided a rather
poor TSP solution, finding only locally optimal paths for problems of touring
more than a few cities, the approach supported the use of more successful ``neural"
solutions to simpler problems. In the
associative memory problem, one simply seeks the closest local optimum; the
issue of avoiding such solutions in favor of a global optimum does not
arise. That a neural solution to the TSP exists at all still speaks to the
power of the neuromorphic approach.

In more recent years the suggestion that the neuronal units of the old-style
networks ought to be replaced by oscillating dynamical systems has become
popular\cite{Chua93}. The oscillations are analogous to spike
trains in real neurons.  The resulting cellular neural networks (CNNs) arguably resemble
biological systems more closely than their older counterparts, in which a
``neuron" exists in a fixed state and outputs a pre-defined, approximately 
binary function of a  collection of inputs. Attempts to capture the properties
of biological systems in simple models, such as 
the Grossberg-Mingolla model of early vision\cite{Grossb}, indeed depend on 
neuronal dynamics.
Further, in the realm of practical application, CNN architectures lend 
themselves to implementation in analog hardware.

A CNN generalization of the Hopfield associative memory network has
recently been proposed\cite {Nish04a,Nish04b}. Some effort was required to
stabilize the desired retrieval state, as compared to the original
Hopfield network. The resulting network performed almost as well
as its fixed-state counterpart, a finding that was taken to imply a broad 
utility for the CNN paradigm.

Relationships between oscillating ``cellular" neurons offer
richer possibilities for representation of solutions and of intermediate
computational steps. Synchronization of two or more
neurons (i.e. phase-locking) was the defining relationship in the proposed
CNN associative memory architecture. More generally, synchronization offers a 
representation of binding that may be useful for image
segmentation and other combinatorial problems\cite{TerWang}.  Such applications are
in line with the suggested role of synchronization in perceptual grouping in
biological systems\cite{Strog,Grossb2,Gray}. We show here that
synchronization in a CNN provides a particularly natural representation of
the TSP, yielding a solution that generalizes the older Hopfield method.
In future work, we will examine the use of synchronization, as an intrinsically
loose form of binding, to avoid the problem of local optima.

We first summarize the original Hopfield solution in the next section, and then
describe the CNN generalization in Section 3.  Results are presented in Section
4, where the essential role of stochasticity in a simulated annealing scheme
is also demonstrated. The stochastic version leads to a suggestion in
the concluding section that the instability that was seen as an obstacle
to the design of CNN's for associative memory may be beneficial in CNN's
designed for optimization.

\section{Background: The Hopfield TSP solution}

The Hopfield-Tank network is a traditional
neural network, fully interconnected, with fixed weights chosen so that the
globally optimal state corresponds to a shortest-distance ``tour" among
a collection of $n$ cities with a pre-specified distance for each pair of cities.
In this original representation, one considers
an $n\times n$ matrix of binary values, where the rows correspond to cities and
the columns correspond to slots, 1 through $n$, in the tour schedule. A tour
is any pattern of 0's and 1's, such that there is exacly one 1 in each row (exactly 
one city visited at a time) and one 1 in each column (each city visited exactly 
once). For instance the tour depicted in Fig. \ref{figHopf}a is ECABD. For the
5-city problem, there is 
a 10-fold degeneracy in optimal patterns (shortest-distance cyclic tours)    
due to arbitrariness in the selection of the starting city and the direction
of the tour.  The travelling salesman problem, in this representation or any
other, is difficult because of the multiplicity of local optima.

The Hopfield and Tank network converges in a few neural time constants for
small $n$, but the required time grows rapidly with $n$ and sub-optimal
solutions are typically found. The nonlinear response function of each
unit is key to the operation of the network.  Flexibility comes from
embedding a discrete problem in a continuous decision space. That is,
each unit $i$ generates a nearly, but not quite binary output as a sigmoid
function $V(u_i)=\frac{1}{2}(1+\tanh(u_i/u_o))$ of a continuous valued 
input $u_i$. Consider a general,
fully interconnected network of $N$ units (here $N=n^2$), where each unit is
updated according to:
\begin{equation}
\label{HopODE}
\frac{du_i}{dt}=-u_i/\tau + \sum_{j=1}^N T_{ij}V_j + I_i
\end{equation}
where $T_{ij}$ is a symmetric matrix defining the connections among
the units in the network, the $I_i$ are biases, and $\tau$ is a decay
time. It can be shown that an energy function
\begin{equation}
\label{energy}
E= -\frac{1}{2} \sum_{i=1}^N \sum_{j =1}^N T_{ij}V_i V_j - \sum_{i=1}^N V_i I_i
\end{equation}
is locally minimized, by computing the time derivative $dE/dt$ and noting that
 the resulting form is negative semi-definite and that
$E$ is bounded below. In the limit of high gain (small $u_o$), for which
the sigmoid function is infinitely steep, the minima only occur when each 
$V_i$ is $0$ or $1$,
i.e. at the corners of the $N$-dimensional hypercube $ [0,1]^N$ that is
the state space of the network described in terms of the variables $V_i$.

The class of optimization problems that can be formulated in this manner
is quite large, as illustrated by the Travelling Salesman Problem, which
is {\it a priori} different in form. The TSP energy is
\begin{eqnarray}
\label{TSPenergy}
E=&&\frac {A}{2} \sum_X \sum_i \sum_{j\ne i} V_{Xi} V_{Xj} 
 +\frac {B}{2} \sum_i \sum_X \sum_{Y\ne X} V_{Xi} V_{Yi} \nonumber \\
 &&+\frac {C}{2}\left[ (\sum_X \sum_i V_{Xi}) - n \right]  \nonumber \\
 &&+\frac {D}{2} \sum_X \sum_{Y\ne X} \sum_i d_{XY} V_{Xi}(V_{Y,i+1}+V_{Y,i-1})
\end{eqnarray}
where the summation indices $X$ and $Y$ range over the $n$ cities, and the
indices $i$ and $j$ range over the $n$ slots in the schedule.
The first term in (\ref{TSPenergy})  inhibits multiple 1's in each row, the second term
inhibits multiple 1's in each column, the third term tends to force the total
number of 1's in the matrix to be exactly $n$, and the last term minimizes
the total distance associated with any pattern of 0's and 1's that defines
a valid tour.

The network method of optimization succeeds largely because difficult decisions
between similar tours can effectively be postponed until the end of a
calculation. In the interim, non-tour patterns (corresponding to multiple
visits to a city or visits to more than one city at a time) can be regarded as 
representing alternative choices simultaneously. Choices can thus be winowed
incrementally toward a true optimum. Nonetheless, the network tends to converge
to local optima, rather than the global optimum, especially as $n$ becomes large.

It was found that global optimization performance could be greatly improved by
a ``simulated annealing" technique in which a stochastic component of gradually 
decreasing amplitude is added to the dynamics.
In the
original Hopfield annealing scheme, the variables $V_i$ were reinterpreted as
expectation values of a statistical distribution of ``spin states", and 
a lowering of ``temperature" was effected by increasing the ``gain" in
the sigmoid function that determines $V_i$, i.e. decreasing $u_o$. The 
stochastic component can also be represented directly as a noise term
in the dynamical equations.

\section{A `cellular' generalization of the Hopfield solution}

\begin{figure}
\begin{minipage}{3in}
  a)\resizebox{1.0\textwidth}{!}{\includegraphics{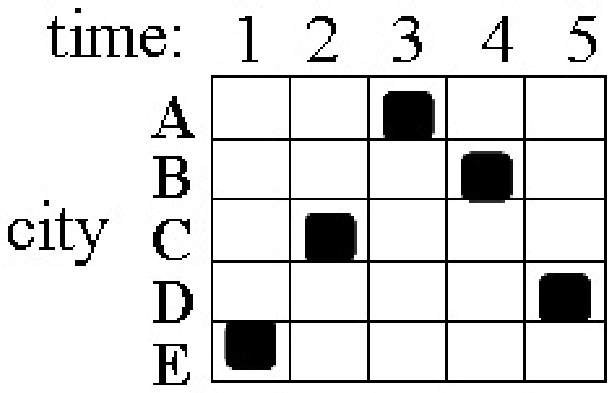}} \\
  b)\resizebox{.8\textwidth}{!}{\includegraphics{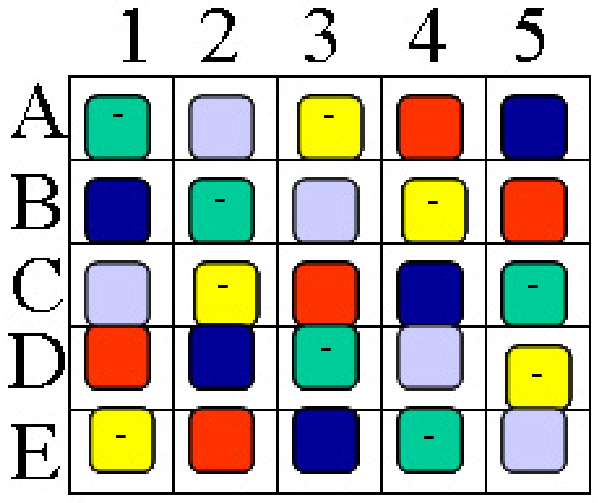}}
\end{minipage} 
\caption{In the original Hopfield representation (a), a tour (here ECABD) is 
 specified by the collection of units that are ``on", provided there is only
 one ``on" unit in each row and only one in each column. In the CNN 
 representation (b) equivalent cyclic permutations of the same tour are simultaneously
 represented as sets of 5 units that are synchronized, with an analogous
 proviso. Units with the same relative phase are shown in the same color.
 (Dash marks distinguish ambiguous colors in the grayscale version.)}
\label{figHopf}
\end{figure}

The representation considered here is an $n \times n$ array of coupled periodic
oscillators. A tour is specified by a synchronization pattern in which all oscillators
in each row and each column are desynchronized, and such that for each oscillator in each
column (row), there is exactly one oscillator in each other column (row) that is synchronized
with it.  In Fig. \ref{figHopf}b, the same tour ECABD is depicted in the new representation,
simultaneously with equivalant tours given by cyclic permutations of cities: CABDE, ABDEC,
etc.  There is now only a two-fold degeneracy in optimal patterns, due to arbitrariness in
direction.

To solve the travelling salesman problem in the representation we have described,
let each
oscillator be given by a complex number $z_{ij}$ ($i=1,2,\dots n \;\;\,j=A,B,\dots X_n$) 
that contains  
both a phase $\arg (z_{ij})$ and an amplitude $|z_{ij}|$. Assume all oscillators
have the same frequency $\omega$ and make the replacement 
$z_{ij} \rightarrow \exp(-i\omega t) z_{ij}$, so that only the relative
phases are represented in the complex quantities $z_{ij}$.   The Lyapunov function 
we seek to minimize is:
\begin{eqnarray}
\label{HopL}
L&=&   A\sum_{ij}(|z_{ij}|^2 -1)^2 
   + B \sum_{ij}\left|\left(\frac{z_{ij}}{|z_{ij}|}\right)^n-1\right|^2 \nonumber \\
   &&-C \sum_i \sum_{jj^\prime} 
                \left|\frac{z_{ij}}{|z_{ij}|} 
          -\frac{z_{ij^\prime}}{|z_{ij^\prime}|}\right|^2   \nonumber \\
   &&-D \sum_j \sum_{ii^\prime} 
                \left|\frac{z_{ij}}{|z_{ij}|}  
                            -\frac{z_{i^\prime j}}{|z_{i^\prime j}|}\right|^2 \nonumber \\
   &&+E \sum_i \sum_j \sum_{j^\prime} d_{jj^\prime}
    {\rm Re}\left(\frac{z_{ij}}{|z_{ij}|}\frac{z^*_{i+1,j^\prime}}{|z_{i+1,j^\prime}|}\right)
\end{eqnarray}
where asterisks denote complex conjugates, and we close the tours by defining
$z_{n+1,j}\equiv z_{1j}$.
The first term, with coefficient $A$, tends to force $|z_{ij}|=1$. The second term,
with coefficient $B$ tends to force each $z_{ij}$ to one of $n$ phase states, 
corresponding to the $n$th roots of unity. The terms with coefficients $C$ and
$D$ penalize for synchronization within each row and within each column,
respectively.
  The last term,
expressed in terms of distances $d_{jj^\prime}$ between
cities $j$ and $j^\prime$, tends to a minimum when each partial sum 
$\sum^{sync}$ over synchronized
oscillators at $i,j$ and $i+1,j^\prime$ (i.e. for which 
$\arg (z_{ij}) = \arg (z_{i+1,j^\prime})$), 
$\sum_{ijj^\prime}^{sync} d_{jj^\prime}$ is minimized,
that is when the specified tour has shortest distance,
since it is expected that such partial sums will dominate the total sum.

If the oscillators are governed by equations
\begin{equation}
\label{Hopz}
{\dot z}_{ij} = - \frac{\partial L}{\partial z^*_{ij}}, \hspace{.3in}
{\dot z}^*_{ij} = - \frac{\partial L}{\partial z_{ij}}
\end{equation}
following Hoppensteadt\cite{Hoppen96},
then the derivative of the Lyapunov function 
$\dot L=\sum_{ij}[(\partial L/\partial z_{ij})\;\dot z_{ij} 
   + (\partial L/\partial z^*_{ij})\;\dot z^*_{ij}] 
          = -2 \sum_{ij} |\dot z_{ij}|^2 \le 0$, so that the cost
function is monotonically decreasing and must reach at least a local minimum,
since $L$ can be easily seen to be bounded below.

For sufficiently large $B$, the units are each forced to one of the $n$
allowed phase states. For the restricted configurations thus defined, it is
easily seen that the sum of the terms
in (\ref{HopL}) with coefficients $C$ and $D$ is minimized when the $n$ phases
in any column or any row are all different. Such configurations define a set
of $n$ tours, some or all of which may be equivalent via cyclic permutation.
Each such tour $j(i)$ defines a set of units with the same relative phases,
giving a contribution to the distance term in (\ref{HopL}) equal to
$E \sum_i  d_{j(i),j(i+1)}
 {\rm Re}(\frac{z_{i,j(i)}}{|z_{i,j(i)}|}\frac{z^*_{i+1,j(i+1)}}{|z_{i+1,j(i+1)}|})
 =  E \sum_i  d_{j(i),j(i+1)}      $
since $\arg(z_{i,j(i)})=\arg(z_{i+1,j(i+1)})$ for synchronized units.
Each of the $n$ such contributions is minimized when each tour length 
$\sum_i  d_{j(i),j(i+1)}$ is minimized. Additional contributions from cross
terms between desynchronized units can be neglected in a stationary phase
approximation.  Coefficients $A,B,C,D,E$ can then be
chosen so that the various optimization criteria, forcing select phase states,
forcing tour configurations, and minimizing distance, can indeed be
satisfied simultaneously. A formal proof is given in
Appendix A.

\section{Results}
\subsection{The n-phase CNN}
\begin{figure}[b]
\resizebox{.35\textwidth}{!}{\includegraphics{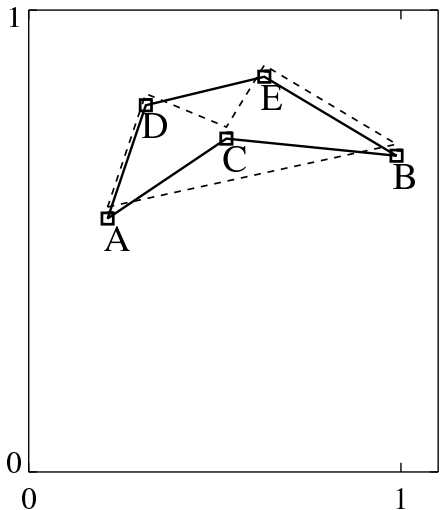}}
\caption{Map, to scale, of the five cities defining the distances $d_{jj^\prime}$ used for all 5-city TSP 
simulations, showing the shortest tour, ACBED (solid line), and the second-shortest
tour, ABECD (dashed line).\cite{AImag}}
\label{figdist}
\end{figure}
The dynamical equations derived from (\ref{HopL}) and (\ref{Hopz}) are:
\begin{eqnarray}
\label{dzdt}
\frac{dz_{ij}}{dt}&=&  -2A (|z_{ij}|^2 -1) z_{ij}  \nonumber \\
   &&- \frac{n}{2} B \left[\left(\frac{z_{ij}}{|z_{ij}|}\right)^n
        - \left(\frac{z^*_{ij}}{|z_{ij}|}\right)^n \right] 
     * \frac{1}{z^*_{ij}} \nonumber \\
   &&+ \frac{1}{2}C \sum_{j^\prime} 
                \left[\frac{z_{ij}}{z^*_{ij}|z_{ij}|} \frac{z^*_{ij^\prime}}{|z_{ij^\prime}|}
          -\frac{1}{|z_{ij}|}\frac{z_{ij^\prime}}{|z_{ij^\prime}|}\right]   \nonumber \\
   &&+ \frac{1}{2}D \sum_{i^\prime} 
                \left[\frac{z_{ij}}{z^*_{ij}|z_{ij}|} \frac{z^*_{i^\prime j}}{|z_{i^\prime j}|}
          -\frac{1}{|z_{ij}|}\frac{z_{i^\prime j}}{|z_{i^\prime j}|}\right]   \nonumber \\
   &&- \frac{1}{4} E  \sum_{j^\prime} d_{jj^\prime}
        \left[\frac{1}{|z_{ij}|}\left(\frac{z_{i+1,j^\prime}}{|z_{i+1,j^\prime}|}
 	                             +\frac{z_{i-1,j^\prime}}{|z_{i-1,j^\prime}|} \right) \right.\nonumber \\
   && \,\,\,\,\,\,  \left.-\frac{z_{ij}}{z^*_{ij}|z_{ij}|}\left(\frac{z^*_{i+1,j^\prime}}{|z_{i+1,j^\prime}|}
 	                             +\frac{z^*_{i-1,j^\prime}}{|z_{i-1,j^\prime}|} \right)\right]
\end{eqnarray}

\begin{figure}[b]
a) \resizebox{.30\textwidth}{!}{\includegraphics{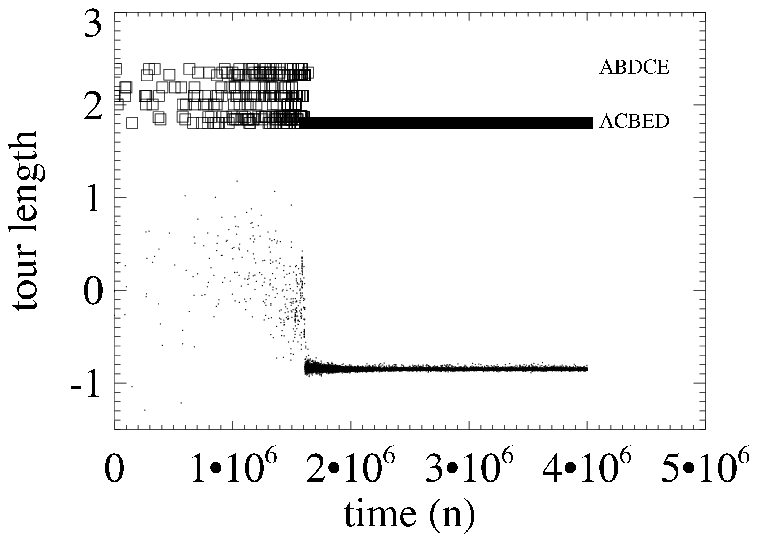}}
b) \resizebox{.30\textwidth}{!}{\includegraphics{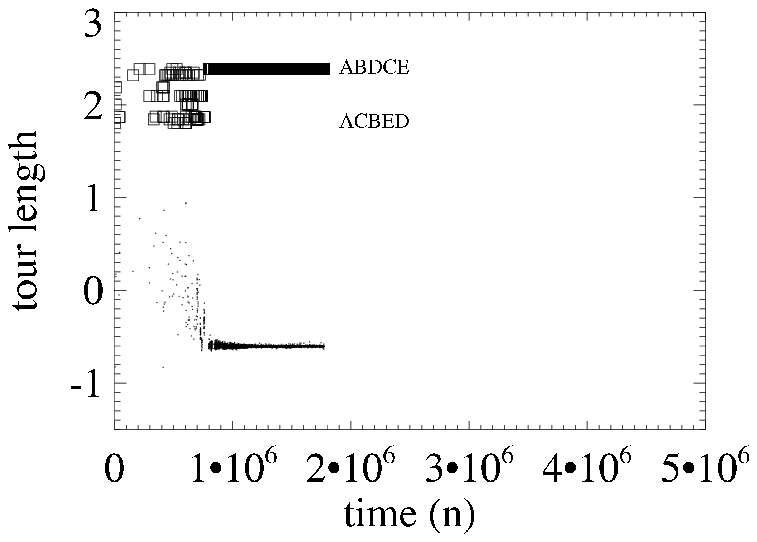}}
c) \resizebox{.30\textwidth}{!}{\includegraphics{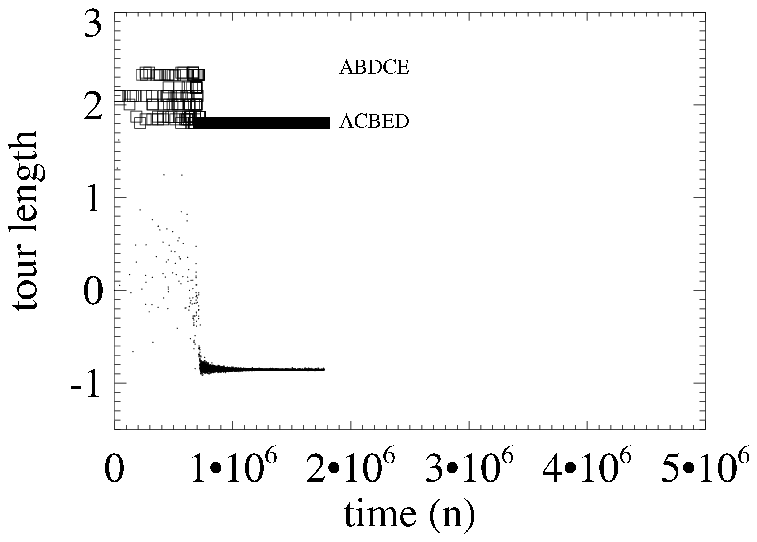}}
\caption{Network histories for three runs of the cellular neural network with parameters and
annealing schedules as in Table \ref{tabphasestoch}. Runs shown are for slow
annealing (a) and fast annealing (b,c), with randomly chosen initial conditions.
Only states (squares) corresponding to tours are
plotted, with tour lengths as indicated. For each such state, a tour length is 
also estimated (dots) as one-fifth the value
of the distance-minimizing part of the Lyapunov function 
(\ref{HopL}), i.e. $(1/5)\sum_i \sum_j \sum_{j^\prime} d_{jj^\prime}
{\rm
Re}\left(\frac{z_{ij}}{|z_{ij}|}\frac{z^*_{i+1,j^\prime}}{|z_{i+1,j^\prime}|}\right)$,
that gives an average over the choice of initial city, and
would be equal to the actual tour length,
if cross-terms for which $\arg(z_{i,j})\ne \arg(z_{i+1,j^\prime})$ could be neglected 
in the sum over $j$ and $j^\prime$ . }
\label{fignethist}
\end{figure}

A network governed by (\ref{dzdt}) almost never reaches
a global minimum, or even a state defining a tour. For the set of pre-specified distances
computed from the map in Fig. \ref{figdist}, the network reached a phase state 
shown in Table \ref{tabphase}, in a typical run.  
However, if gaussian
noise is added to the dynamics at periodic intervals:
\begin{equation}
\label{Hopznoise}
z(t+\epsilon)=z(t)\xi(t)
\end{equation}
whenever $t=k\tau$ for some integer $k$ and
where $\xi$ is a  random complex number, 
then the system can be made to attain the desired state, provided that the amplitude 
of the noise in the phase of $\xi$
(an analog ``temperature")
is decreased slowly
enough. We used $\xi(k\tau)=\rho \exp(i \Theta_k)$, where the magnitude
$\rho$ is uniformly distributed between $0.7$ and $1.3$, and the random variable 
$\Theta_k$ in the phase is a unit
gaussian deviate with standard deviation $\sigma^\Theta_k$,
that steadily decreases over time, i.e. $\sigma^\Theta_k = \alpha \sigma^\Theta_{k-1}$, starting
from a value $\sigma_0=4$ that effectively randomizes the initial phase perturbations.  Resulting 
states such as the one shown in 
Table \ref{tabphasestoch} are indeed tours. Typical convergence histories are shown in Fig. 
\ref{fignethist}. The tour state in Table \ref{tabphasestoch} corresponds to
the shortest-distance tour among the 12 distinct tours shown in Table \ref{tourtable},
 for the table of distances that was used to 
construct
the network.  This direct simulated annealing method is effective in selecting one of a 
small number of global
optima (precisely $2\times 5!=240$, since there are $5!$ assignments of phase states to synchronized
subsets, and two overall choices of direction) out of $5^{25}$ network states.
Locally optimal states are sometimes selected for a given annealing schedule, but
the procedure appears to converge to a perfect one as the annealing rate is lowered.
That is, 
as shown in Table \ref{tourtable}, the network appears more likely to converge to the shortest-distance
tour as the analog ``temperature" is lowered more slowly. 

The highly irregular convergence that is seen is to be expected as the system randomly jumps among
basins of attraction corresponding to different local optima as the temperature is lowered. Only some
of these optima correspond to tours. The more slowly the temperature is lowered, the greater the chance
that the system will reach a ``deep" global optimum, corresponding to a shortest-distance tour, and have 
insufficient energy to escape at that point in the process. The convergence pattern is typical of simulated
annealing, as reported for instance by Kawabe {\it et al.} \cite{irrconv}.

\begin{table}[t]
\caption{Relative phases (from $-\pi$ to $\pi$) of oscillators in the $5\times
5$ CNN representation of the 5-city travelling salesman problem, after
reaching a steady state, with coefficients $A=0.5$, $B=0.08$, $C=D=4.0$, 
$E=0.4$. The dynamical equations (\ref{dzdt}) were integrated using a finite
difference scheme with time step $\Delta t= 0.01$.  No tours are evident.} 
\label{tabphase}
\begin{tabular}{cr|ccccc}
\hline
     & & \multicolumn{5}{c}{slot in schedule} \\
     & &  1    & 2 & 3  & 4 & 5    \\
     \hline
        &A      & -2.470  & -2.187  & -0.144   & 1.284   & 1.286 \\
        &B      &  2.496  & -1.221  &  1.404   &-2.484   &-0.012 \\
{city  }&C      &  1.240  &  2.643  & -2.619   & 0.009   &-1.221 \\
        &D      &  0.004  &  1.088  & -2.637   &-1.271   & 2.533 \\
        &E      & -1.292  &  1.112  &  0.049   & 2.519   &-2.530 \\
\hline
\end{tabular}
\end{table}

\begin{table}[t]
\caption{Relative phases as in Table \ref{tabphase}, but with noise of
steadily decreasing amplitude included in the oscillator dynamics
(\ref{Hopznoise}), using a noise reduction factor $\alpha=0.9999993$ every five
time steps, i.e. at time
intervals $\tau=0.05=5\Delta t$. The tour found, ADEBC, is seen by focussing attention
on oscillators with relative phase $-2\pi/5\approx -1.28$ (bold). Permutations
of the same tour are represented by other collections of oscillators with
approximately equal relative phases, e.g. those with phase $\approx 0$, defining
the equivalent tour CADEB (italics).  The tour class found is indeed the shortest-distance
tour class for the pre-specified table of distances ($d_{jj^\prime}$) computed
from Fig. \ref{figdist}.}
\label{tabphasestoch}
\begin{tabular}{cr|ccccc}
\hline
     & & \multicolumn{5}{c}{slot in schedule} \\
     & &  1    & 2 & 3  & 4 & 5    \\
     \hline
        &A      & {\bf -1.282}   & {\it 0.036}   & 1.240         & 2.494        & -2.521 \\
        &B      &  1.233         & 2.485         &-2.500         &{\bf -1.320}  & {\it -0.016} \\
{city  }&C      & {\it  0.034}   & 1.280         & 2.524         &-2.537        & {\bf -1.259} \\
        &D      & -2.498         &{\bf -1.293}   &{\it -0.013}   & 1.195        &  2.502 \\
        &E      &  2.493         &-2.423         &{\bf -1.190}   & {\it 0.046}  &  1.285 \\
\hline
\end{tabular}
\end{table}

Varying the different coefficients affects the network in a way similar to the variation of corresponding
coefficients in the original fixed-state network. The largest effect comes from varying $E$, which multiplies
the distance term, but increasing $E$ tends to force the network toward non-tour states. Increasing $E$ fourfold
to $E=1.6$ tended to favor shorter-distance tours among the runs that converged to tours, as shown in the table, but
most runs converged to non-tour states. No tours were
obtained in 20 runs with a tenfold increase of the coefficient to $E=4.0$.

The statistical significance of the selection of tour states over non-tour
states 
is obvious. The observed tendency of the network to select shortest-distance
tours is weaker, but still statistically significant. Formally, the
inverse correlation of the number of occurrances of each tour class with tour
length is significant at the 95\% level for all three cases shown, and at the
99\% level for the case of slow annealing. The tendency of the network
to specifically select the shortest tour ACBED, among the twelve tour classes,
is significant at the 97\% level for slow annealing, and at the 99.9\%
level for fast annealing with $E=1.6$. Additional details of the significance
calculations are provided in Appendix B.  The observed trends for slower
annealing and for increased $E$ add to the significance.

\begin{figure}[b]
a) \resizebox{.40\textwidth}{!}{\includegraphics{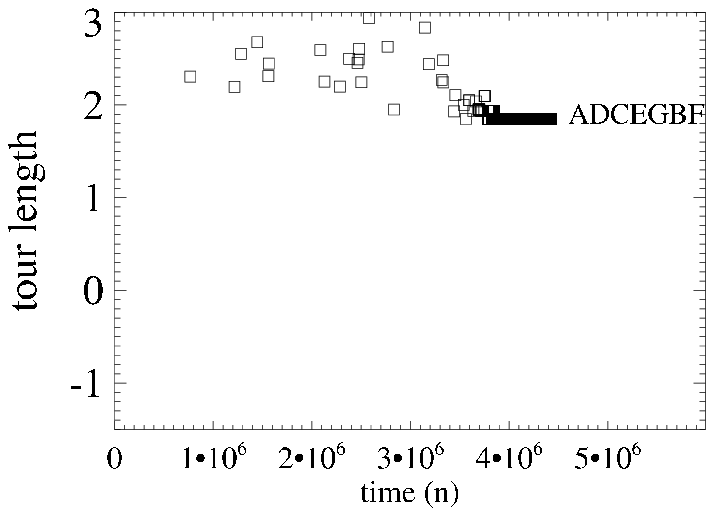}}\hspace{.5in}
b) \resizebox{.30\textwidth}{!}{\includegraphics{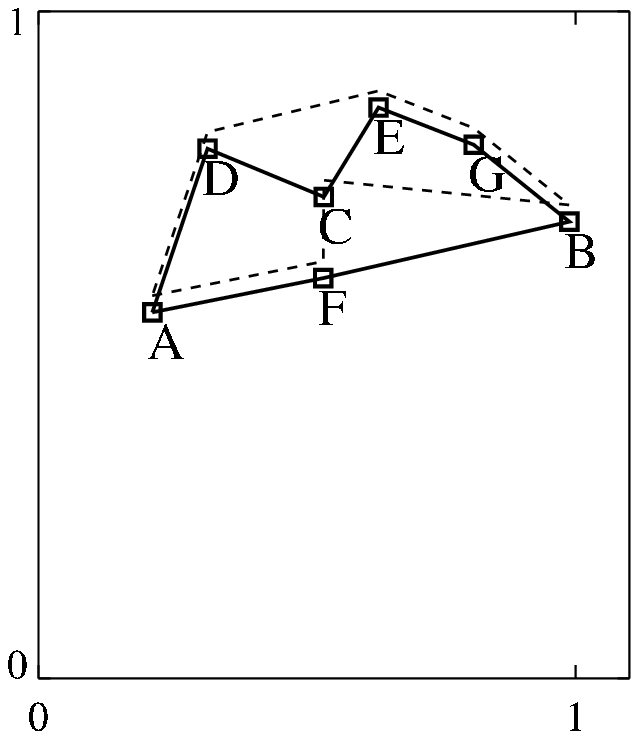}}
\caption{Network history (a) for a run of the cellular neural network for 7 cities
plotted as in Fig. \ref{fignethist}. Network parameters are as in the 5-city case, but
with $C=D=16.0$ and annealing schedule given by $\alpha=0.99999995$, $\tau=0.02=2\Delta t$. 
In two out of three runs with the city map (b), the network converged to tour states: the optimal
tour ADCEGBF (solid) and a suboptimal tour AFCBGED (dashed).}
\label{fighop7}
\end{figure}

\begin{figure}[t]
a) \resizebox{.40\textwidth}{!}{\includegraphics{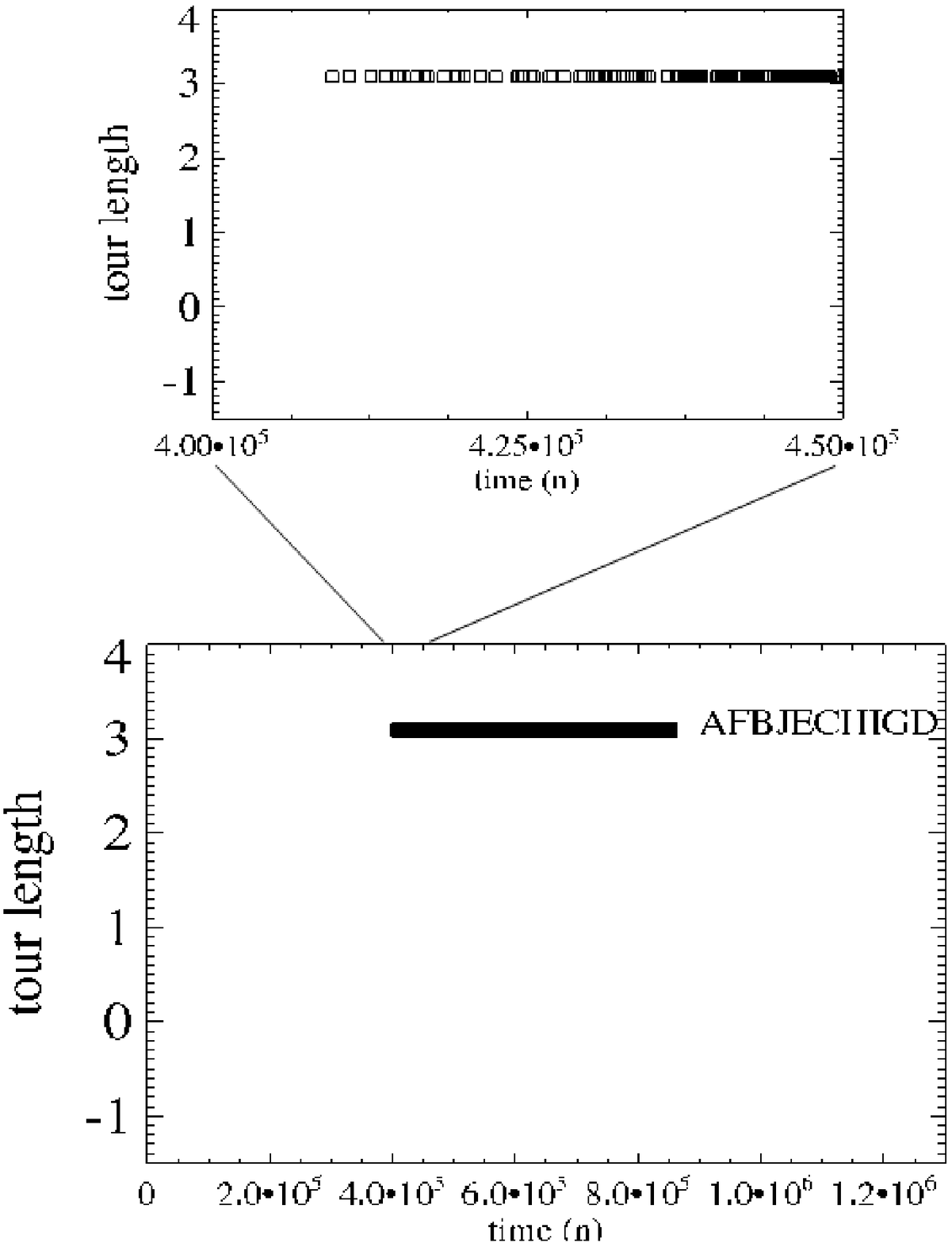}}\hspace{.5in}
b) \resizebox{.30\textwidth}{!}{\includegraphics{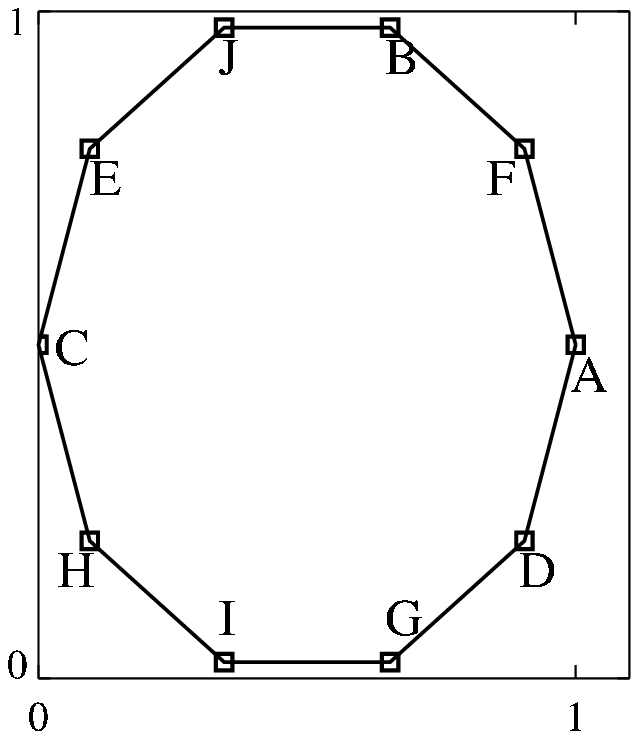}}
\caption{Network history (a) for a run of the cellular neural network for 10 cities
plotted as in Fig. \ref{fignethist}. Network parameters are as in the 5-city case, with
with a ``slow annealing" schedule ($\alpha=0.9999998$, $\tau=0.02=2\Delta t$). 
With cities placed on a circle, with uniform spacing, in arbitrary order (b), the network converged to the
optimal tour. }
\label{fighop10}
\end{figure}

For a larger number of cities $n$, preliminary experiments indicate that the architecture
continues to select tour states over non-tour states and prefers shortest-distance tours.
The convergence time increases, but is strongly dependent on the configuration of cities
and the associated ``energy" landscape.  Typical results for an $n=7$ problem are shown in
Fig. \ref{fighop7}.  Parameters are the same as in the $n=5$ example, except that the coefficients
on the terms that force de-synchroniation within the separate rows and columns had to be increased
fourfold and the annealing schedule further slowed. In two out of three runs, the network converged to a tour state, in about twice the number
of iterations required for $n=5$. For $n=10$, with the simple configuration shown in Fig. \ref{fighop10}b, and 
with no changes in parameters as compared to $n=5$,  the network converged to a globally optimal tour state 
even more quickly than in the $n=5$ example (Fig. \ref{fighop10}a). For other 10-city configurations, 
the network did not converge to
any tour or non-tour state, even with vanishing noise level, with the same fixed-step numerical scheme used
in the other examples. This is to be attributed to the complexity of the landscape and the large number
of closely spaced local optima. An analog implementation, the ultimate target of the initial digital
investigation, would obviate the numerical issues.

Distance minimization is achieved in all cases because of the validity of the stationary
phase approximation, according to which the minimum of the term in
the Lyapunov function with coefficient $E$ is also the minimum of the partial
sum without cross-terms.  Convergence histories of the total sum in this term for the 5-city example are plotted
with the histories of the corresponding partial sums in Fig. \ref{fignethist}.
For the case shown in Fig. \ref{fignethist}b, where a non-optimal
tour is selected, the total sum is also seen to be slightly higher than
the minimal value attained in the other two panels. 
The partial and total sums are widely separated, indicating that the cross-terms do
not cancel, but sum to effectively random values that have no average effect
on the minimization of the partial sums over synchronized units.

\begin{table}
\caption{Tour lengths for the twelve tour classes in the 5-city problem, with
the number of times each class occurs as the final state in 20 runs
of the CNN architecture using a fast annealing schedule ($\alpha=0.999999$, $\tau=0.05=5\Delta t$),
in 20 runs with the same fast annealing schedule but with the weight $E$ of the
distance-minimizing term increased fourfold, and in 20 runs with a slow annealing schedule 
($\alpha=0.9999998$, $\tau=0.02=2\Delta t$). Other parameters are as in
Table \ref{tabphase}. Synchronization, as used to 
define tour patterns, is deemed to occur when the relative phases of two
or more units differ by less than $0.6$.}
\label{tourtable}
\begin{tabular}{c|c|c|c|c}
\hline
 final state  & tour length & \multicolumn{3}{c} {number of occurrences} \\
 \cline{3-5}
tour class  & & \multicolumn{2}{c}{fast annealing} & slow annealing \\
\cline{3-4}
& & $E=0.4$ & $E=1.6$ & \\
\hline
ACBED & 1.806 & 1 & 4 & 4 \\
ABECD & 1.845 & 4 & 1 & 3 \\
ADCBE & 1.868 & 4 &   & 1 \\
ACEBD & 1.874 & 1 & 1 & 3 \\
ABCED & 2.005 &   &   & 1 \\
ADBCE & 2.096 & 2 &   & 1 \\
ABEDC & 2.098 & 1 &   & 1 \\
ACDBE & 2.189 &   &   &   \\
ABCDE & 2.320 &   &   &   \\
ABDEC & 2.326 &   &   &   \\
ACBDE & 2.349 & 1 &   &   \\
ABDCE & 2.388 &   &   &   \\
non-tour &    & 6 & 14& 6 \\
\hline
\end{tabular}   
\end{table}

\subsection{The CNN with unrestricted phases}
The second term in (\ref{HopL}), restricting the phase of
each oscillator to one of $n$ values, can be dropped, but the temperature must be
lowered even more slowly to attain a globally optimal pattern.  Adding noise
at every time step, at a level that is reduced by the factor $\alpha=1-1/10^7$ 
after each time
step, we obtained results for a single long run that are displayed in Table \ref{taballphase}.  The shortest-distance
tour was indeed selected, but the global optimality of such states for the general problem
remains to be proved.
\begin{table}[t]
\caption{Relative phases as in Table \ref{tabphase}, but for a network with
$B=0$ in which the phases are unrestricted, and with very slow annealing, given by
$\alpha=0.9999999$ and $\tau=0.01=\Delta t$  in
(\ref{Hopznoise}). The other CNN coefficients are $A=0.5$, $C=D=20.0$, and 
$E=0.12$. 
After a very large number of  iterations, the network converged to the shortest 
tour.}
\begin{tabular}{cr|ccccc}
\hline
     & & \multicolumn{5}{c}{slot in schedule} \\
     & &  1    & 2 & 3  & 4 & 5    \\
     \hline
      &A      & {\bf 2.316}   &{\it -2.583}   &-1.209        &-0.151        &  1.289 \\
      &B      & -1.607        & 0.129         & 1.237        & {\bf 2.335}  &{\it -2.286} \\
{city}&C      & {\it -2.878}  & -1.291        & 0.321        & 1.227        &{\bf 2.372} \\
      &D      &  0.929        &{\bf 2.600}    &{\it 3.032}   &-1.482        & 0.461 \\
      &E      &  0.392        & 1.077         &{\bf 2.344}   &{\it -2.610}  &-1.388 \\
\hline
\end{tabular}
\label{taballphase}
\end{table}

\section{Conclusions and Proposed Extensions}

Synchronous firing of neurons, sometimes widely separated in the brain, is
a ubiquitous phenomenon that has been advanced as an explanation for
perceptual grouping\cite{Gray,Schechter,vdM1}, 
motor control\cite{Vaadia,Kopell}, and consciousness\cite{vdMalsburg, Rodriguez,
KochGreen}.
Information in the brain may generally be coded in synchronization patterns.  
The dynamical
evolution of such patterns would then define mental processing. One may inquire as
to the utility of synchronization patterns for representation of objective states.

The main
result of this paper is that the
evolution of synchronization patterns can be organized so as to
satisfy a complex optimization principle.  Indeed, the 5-city architecture is
effective in selecting $2\times 5!=240$ shortest-distance tours out of $5^{25}$
possible states. The original Hopfield architecture, by contrast, would select $2\times
5=10$ binary-valued tour states out of $2^{25}$ possibilities. The task of the synchronization
network is more difficult by a factor of 
$\frac{10}{240}\left(\frac{5}{2}\right)^{25}=3.7\times10^8$, a factor which
grows exponentially with the number of cities. The landscape searched by the synchronization network
is correspondingly rougher, requiring shorter time steps in a digital implementation.  A 
large research effort with the original
Hopfield architecture resulted in effective performance only for relatively small $n$, up to
about 100. 
  Preliminary indications are that the synchronization network will be
competitive for simple configurations of cities and may reach at least the original
Hopfield-Tank\cite{HopTank} level ($n=30$) for arbitrary configurations, with improved numerics or in an
analogue implementation. The demonstrated skill is surprising in view of the increased difficulty of
the task in the new representation. Further, the network's sensitivity to the complexity of the map
is suggestive of a human's response to the same problems. Thus the results
support synchronization-based theories of biological information processing.

Freeman\cite{Freeman, FreemanSA} has suggested a fundamental role for chaos in intelligent information
processing. (See also Nara\cite{Nara}.) In Freeman's view, chaos provides a combination of 
sensitivity and stability
that might account for the observed range of mental competence in the face of
unique inputs. The known highly intermittent synchronization of 40Hz oscillations was taken
to support the suggested role for chaos. Information is thought to be coded in spatial
patterns of coherent activity that exist for very brief periods of time, typically
less than 200 milliseconds.

Synchronization networks of the type described in this paper can be naturally extended
to a chaotic framework. If the limit cycle oscillators were to be replaced by chaotic
oscillators, chaos might replace noise in the simulated annealing scheme. (The ``chaotic oscillators" might 
be higher-level constructs formed from groups of units with incommensurate frequencies.) Synchronization
of chaotic oscillators typically degrades via on-off intermittency, wherein the synchronized
phase is interrupted at irregular intervals by bursts of de-synchronization\cite{PST,OS}. 
A stable synchronization manifold, like other
invariant manifolds, typically contains unstable periodic orbits (UPOs) embedded within strange attractors. 
Trajectories that approach such orbits while still off-manifold commonly burst away. That behavior would
resemble the noise-induced jumping among different synchronization patterns in the network of regular
oscillators studied here. 

Bursts
away from local optima might play the role of a ``reset" mechanism
such as the one introduced in Adaptive Resonance Theory\cite{ART} for old-style networks. 
Unlike the situation with associative 
memory networks, continuous reset is here desirable. The dynamical regime of nearly
continuous bursting \cite{Duane96} might be
identified with the biological case, in which the synchronized phase is very short-lived.
Slow Hebbian learning, to adjust connection strengths, might be used for ``permanent"
global optimization, if desired, replacing the gradual lowering of temperature in the
simulated annealing scheme. Such chaotic architectures will be the subject of future
work. The simplified limit cycle version confirms the representational utility of 
synchronized oscillator networks.

{\bf Acknowledgements:}  The author thanks Frank Hoppensteadt for encouraging the publication
of the preliminary, regular-oscillator version of the proposed CNN architecture. A fraction of
this work was supported by NSF Grants 0327929 and 0838235.  The National Center for Atmospheric
Research is sponsored by the National Science Foundation.

\newpage
\appendix
\noindent{\bf APPENDIX A: Proof of the optimality of shortest-distance tours for the n-phase 
synchronization network}

Without loss of generality, consider the TSP for a table of distances expressed in units
such that $d_{ij} \le 1$ for all $i,j$. The Lyapunov fuction, previously expressed as
(\ref{HopL}) is:
\begin{eqnarray}
\label{HopL1}
L&=&   A\sum_{ij}(|z_{ij}|^2 -1)^2 
   + B \sum_{ij}\left|\left(\frac{z_{ij}}{|z_{ij}|}\right)^n-1\right|^2 \nonumber \\
   &&-C \sum_i \sum_{jj^\prime} 
                \left|\frac{z_{ij}}{|z_{ij}|} 
          -\frac{z_{ij^\prime}}{|z_{ij^\prime}|}\right|^2   \nonumber \\
   &&-D \sum_j \sum_{ii^\prime} 
                \left|\frac{z_{ij}}{|z_{ij}|} 
                            -\frac{z_{i^\prime j}}{|z_{i^\prime j}|}\right|^2 \nonumber \\
   &&+E \sum_i \sum_j \sum_{j^\prime} d_{jj^\prime}
    {\rm Re}\left(\frac{z_{ij}}{|z_{ij}|}
      \frac{z^*_{i+1,j^\prime}}{|z_{i+1,j^\prime}|}\right)
\end{eqnarray}
The term 
\begin{equation}
\label{eqD}
{\cal D}\equiv \sum_i \sum_j \sum_{j^\prime} d_{jj^\prime}
    {\rm Re}\left(\frac{z_{ij}}{|z_{ij}|}
      \frac{z^*_{i+1,j^\prime}}{|z_{i+1,j^\prime}|}\right) 
\end{equation}
is intended to minimize total tour length. Consider strict tour configurations
for which each $z_{ij}$ is one of the $n$th roots of unity,
i.e. $z_{ij}=\exp(2\pi i m_{ij}/n)$ for some integer $m_{ij}$, 
$0 \le m_{ij} < n$, and the integers ${m_{ij}}$  are all different within
each column and within each row.  That is, we allow 
$m_{ij}=m_{i^\prime j^\prime}$ only
if $i=i^\prime$ and $j=j^\prime$ or if $i\ne i^\prime$ and $j\ne j^\prime$.
Among such strict tour configurations, let ${m_{ij}^0}$ be chosen so as to 
minimize ${\cal D}$.

In a stationary phase approximation, one could relate 
${\cal D}$ to a sum $\hat{\cal D}$ without the cross terms: 
\begin{equation}
\label{eqDhat}
\hat{\cal D}\equiv \sum_i \sum_{\{j j^\prime| m_{i+1,j^\prime} = m_{ij}\}} d_{jj^\prime}
    {\rm Re}\left(\frac{z_{ij}}{|z_{ij}|}
      \frac{z^*_{i+1,j^\prime}}{|z_{i+1,j^\prime}|}\right) \nonumber
\end{equation}
We do not require equality or proportionality of ${\cal D}$ and $\hat{\cal D}$, but make a
weaker assumption that tour configurations that minimize ${\cal D}$ also
minimize $\hat {\cal D}$, i.e.
\begin{equation}
\label{minDhat}
\hat{\cal D}(\{m_{ij}^0\})\le \sum_i \sum_{\{j j^\prime| m_{i+1,j^\prime} =
m_{ij}\}} d_{jj^\prime}
    {\rm Re}\left(\frac{z_{ij}}{|z_{ij}|}
      \frac{z^*_{i+1,j^\prime}}{|z_{i+1,j^\prime}|}\right) 
\end{equation}
for any tour configuration given by $\{z_{ij}=\exp(2\pi i m_{ij}/n)\}$ that is not necessarily
shortest-distance. We regard (\ref{minDhat}) as
empirically validated.

We can then show that the Lyapunov function (\ref{HopL1}) is
appropriate for the TSP by proving the following theorem.

{\it Theorem:}  There are values of the coefficients $A,B,C,D,$ and $E$, in 
the
Lyapunov function $L: {\cal C}^{n^2}\rightarrow {\cal R}$ that is defined in
(\ref{HopL1}), such that the global minima of $L$ occur at states 
$\{z_{ij}:i,j=1\dots n\}$ that correspond to
shortest-distance tours, and only occur at such states, provided that the
assumption (\ref{minDhat}) holds. The required coefficients can be
chosen universally for the n-city problem, except for dependence on the 
difference in the value of ${\cal D}$ (as defined in (\ref{eqD}))
between the shortest and the second-shortest tour configurations.

{\it Proof:}

The first term in (\ref{HopL1}) is a function only of the $n^2$ magnitudes $|z_{ij}|$, while
the remaining terms are functions only of the $n^2$ phases that define the normalized
values $z_{ij}/|z_{ij}|$. Thus the first term, which is non-negative, can be minimized separately, by choosing all $|z_{ij}|=1$, so that the term vanishes.

Note that the third, fourth, and fifth terms
satisfy the respective bounds:
\begin{equation}
\label{Cbound}
0 \ge -C \sum_i \sum_{jj^\prime} 
                \left|\frac{z_{ij}}{|z_{ij}|} 
          -\frac{z_{ij^\prime}}{|z_{ij^\prime}|}\right|^2 \ge -4 n^3 C 
\end{equation}

\begin{equation}
\label{Dbound}
0 \ge -D \sum_j \sum_{ii^\prime} 
                \left|\frac{z_{ij}}{|z_{ij}|} 
       -\frac{z_{i^\prime j}}{|z_{i^\prime j}|}\right|^2 \ge -4 n^3 D
\end{equation}

\begin{equation}
\label{Ebound}
n^3 E  \ge
E \sum_i \sum_j \sum_{j^\prime} d_{jj^\prime}
    {\rm Re}(\frac{z_{ij}}{|z_{ij}|}\frac{z^*_{i+1,j^\prime}}{|z_{i+1,j^\prime}|})
\ge -n^3 E 
\end{equation}

We first show that each normalized value $z_{ij}/|z_{ij}|$ can be made to differ from
one of the $n$th roots of unity by an arbitrarily small amount $\epsilon_{ij}$ when
$L$ is minimized.  Define the deviations $\epsilon_{ij}$ by
\begin{equation}
\label{defeps}
\epsilon_{ij}^{2n} \equiv
\left|\left(\frac{z_{ij}}{|z_{ij}|}\right)^n-1\right|^2
\end{equation}
for a set of values $\{z_{ij}\}$ that minimizes $L$.
If the remaining coefficients $C=D=E=0$, the minima of $L$ occur when each $\epsilon_{ij}=0$.
For nonvanishing $C$,$D$, and/or $E$, no summand in the term with coefficient $B$ can raise
the minimum value by an amount greater than the range of values of the remaining three terms, so $B \epsilon_{ij}^{2n} < 4 n^3 C + 4 n^3 D + 2 n^3 E $, or
\begin{equation}
\label{epsbound}
\epsilon_{ij}^{2n } < (4 n^3 C + 4 n^3 D + 2 n^3 E )/B 
\end{equation}
Thus, by choosing $B$ sufficiently large as compared to $C$,$D$, and $E$, each $\epsilon_{ij}$
can be made as small as desired.

Choose $B$  so that each normalized value $z_{ij}/|z_{ij}|$ is much closer to one particular
$n$th root than to any other, i.e. $\epsilon_{ij} << |1-\exp(2\pi i/n)|$, as will follow
from (\ref{epsbound}) if 
\begin{equation}
\label{Bset}
(4 n^3 C + 4 n^3 D + 2 n^3 E )/B << |1-\exp(2\pi i/n)|^{2n}
\end{equation}
For any given set of deviations $\{\epsilon_{ij}\}_{1\le i,j \le n}$, one is free to
choose one of the $n$ roots arbitrarily for each member of the set. It is easily seen
that the terms with coefficients $C$ and $D$ are minimized when the $n$ values in each
row and each column, respectively, are spaced as widely as possible about the unit circle,
while the term with coefficient $B$ is invariant. Thus if $E=0$, the minima of $L$ occur
when the phases that are approximately $n$th roots of unity are all different in each
row and each column, i.e. for configurations that define  $n$ ``tours". 

If $E$ is sufficiently small, then the minima of $L$ still occur at tour configurations.
Specifically, let $\Delta$ be the minimum difference between the value of a sum over a single
row or column ($\sum_{ii^\prime}$  or $\sum_{jj^\prime}$ in (\ref{HopL1}))  for all phases different 
and the sum for the case of one phase duplication. 
One finds
\begin{equation}
\label{defDelta}
\Delta \equiv |1 - \exp(2\pi i/n)|^2
\end{equation}
Then, using (\ref{Ebound}), the minima of $L$ occur at tours if
\begin{equation}
\label{Esmall}
n^3 E  < C \Delta + O(\epsilon^2)  \hspace{.5in} n^3 E   < D \Delta + O(\epsilon^2)
\end{equation}
where $\epsilon \equiv \max_{ij}\{\epsilon_{ij}\}$.

For strict tour configurations (with all $\epsilon_{ij}=0$), the last term in 
(\ref{HopL1}), $E{\cal D}$, is separately minimized for shortest tours, since
by assumption (\ref{minDhat}), $E{\cal D}$ is minimized for values
$z_{ij}=\exp(2\pi i m^0_{ij}/n)$ that minimize $E \hat{\cal D}$. The latter
can be written as a sum
over $n$ tours, $\kappa =1,\ldots,n$, each tour defined by a set of $n$ units 
with equal phases:
\begin{equation}
\label{toursum}
E\hat{\cal D}
 =E \sum_{\kappa=1}^n d_{j_\kappa(i),j_\kappa(i+1)} 
\end{equation}
where  $\kappa$ labels the tour and $j_\kappa(i)$ is the city visited at time $i$ while on tour $\kappa$.
This sum is minimized if and only if all the $j_\kappa$ are shortest-distance
tours.

For non-strict tour configurations with $\epsilon_{ij} \ne 0$
\begin{equation}
\label{Dapprox}
{\cal D}=\sum_i \sum_j \sum_{j^\prime} d_{jj^\prime}
    {\rm Re}\left(e^{2 \pi i m_{ij}/n} e^{-2 \pi i m_{i+1,j^\prime}/n}\right) 
    + O(\epsilon^2)
\end{equation}
for some $\{m_{ij}\}$.
This sum is minimized for shortest-distance tours provided that the difference 
$\delta \equiv |{\cal D}(\{m_{ij}^0\}) - {\cal D}(\{m_{ij}^1\})|$ in the value of ${\cal D}$ between the 
shortest tour
configuration $\{m_{ij}^0\}$ and the second-shortest tour configuration 
$\{m_{ij}^1\}$  is 
not too small, i.e.
\begin{equation}
\label{mindelta}
\delta >> O(\epsilon^2)
\end{equation}
which is satisfied, according to the bound (\ref{epsbound}), if
\begin{equation}
\label{Bset2}
(4 n^3 C + 4 n^3 D + 2 n^3 E )/B << \delta^n
\end{equation}

For given  $\delta$, we note that the coefficients $B,C,D,$ and $E$ can always 
be chosen so as to satisfy (\ref{Bset}),(\ref{Esmall}), and (\ref{Bset2}), 
completing the proof.

\newpage
\noindent{\bf APPENDIX B: Statistical significance of the results in Table \ref{tourtable}}

The correlation between tour-length $d_i$ 
and number of occurrences $n_i$, as listed in the second and third
columns of Table \ref{tourtable}, respectively,  for the case with fast annealing and $E=0.4$,  is
$r=\sum_{i=1}^{12}d_i n_i/(\sigma_d \sigma_n) =-0.62$. If the data were truly {\it un}correlated, the correlations would
be t-distributed about r=0. In such a distribution with $12-2=10$ degrees of 
freedom,, $r=-0.62$ corresponds to a t-score of $t=2.50$, and the
probability of $r\le -0.62$ is found to be less than 5\%, using a t-distribution
table or online calculator\cite{tcalc}. So the negative
correlation is significant at the 95\% level. 

The significance of the rate of occurrence of the
optimal tour, ACBED, among all tours found, is calculated as follows: In the slow annealing case,
ACBED occurs in 4 instances out of the 14 tour states chosen. If all 
12 possible tour
states were selected with equal probability, the probability of 4 or
more occurrences of the shortest-distance tour would be:
\begin{equation}
\sum_{i\ge 4} 
\left(\begin{array}{c} 14 \\ i \end{array}\right) 
     \left(\frac{1}{12}\right)^i \left(\frac{11}{12}\right)^{14-i}
				    =0.024  \nonumber
\end{equation}
implying significance at the 97\% level.  

Significance levels of the correlations and of the shortest-distance tour selection rates are 
computed similarly for the other 5-city networks, with results as shown in Table \ref{sigtable}. (The fast-annealing
scheme with $E=0.4$ did not select the optimal tour at a significant rate.)   

The trends in Table \ref{tourtable} as $E$ is increased and as annealing is slowed lend additional significance
to the results.

\begin{table}[b]
\caption{Correlation $r$ between length of a tour and frequency of selection of the tour for the
three network configurations represented in Table \ref{tourtable}, with the t-score of each
correlation, the significance level, and the significance level (if any) of the selection
of the shortest-distance tour.}
\label{sigtable}
\begin{tabular}{r|c|c|c}
\hline
     &  \multicolumn{2}{c}{fast annealing} & slow annealing \\
\cline{3-4}
&  $E=0.4$ & $E=1.6$ & \\
\hline
$r$ & -0.62 & -0.59 & -0.87 \\
t-score & 2.50 & 2.30 & 5.58 \\
significance level of distance/frequency correlation& 95\% & 95\%  & 99.9\% \\
significance level of shortest-distance tour selection&  - & 99.9\% & 97\% \\
\hline
\end{tabular}   
\end{table}

\end{document}